\documentclass[aps,onecolumn,showpacs,nofootinbib,showkeys]{revtex4-2}
\usepackage[margin=3.15cm]{geometry}

\RequirePackage[T1]{fontenc}

\usepackage{epsfig,graphicx,amsmath,amssymb,bm}
\RequirePackage{mathptmx}      
\usepackage{mathrsfs}
\usepackage[english]{babel}
\usepackage{amssymb}
\RequirePackage{color}
\usepackage{changes}
\usepackage{slashed}
\usepackage{multirow}
\usepackage{scalerel}
\usepackage{tikz-feynman}
\usepackage{textcomp}
\usepackage{subcaption}
\usepackage[english]{babel}
\usepackage{float}
\RequirePackage{hyperref}
\hypersetup{
    linktocpage,
    colorlinks,
    citecolor=blue,
    filecolor=black,
    linkcolor=blue,
    urlcolor=blue,
}
\usepackage{nccmath}

\begin{document}

\title{Weak semileptonic decays of vector mesons in the NJL model}


\author{M.K. Volkov$^{1}$}\email{volkov@theor.jinr.ru}
\author{A.A. Pivovarov$^{1}$}\email{pivovarov@theor.jinr.ru}
\author{K. Nurlan$^{1,2,3}$}\email{nurlan@theor.jinr.ru}

\affiliation{$^1$ Bogoliubov Laboratory of Theoretical Physics, JINR, 
                 141980 Dubna, Moscow region, Russia \\
                $^2$ The Institute of Nuclear Physics, Almaty, 050032, Kazakhstan}   


\begin{abstract}
The decay widths of $V \to K(\pi) l \bar{\nu}_l $ are calculated within the Nambu--Jona-Lasinio model, where $V=K^*, \rho, \omega, \phi$ and $l=\mu, e$. The results are obtained using the previously fixed model parameters without introducing any arbitrary parameters. The obtained results are considered as predictions due to the absence of experimental data. 

\end{abstract}


\maketitle

\section{Introduction}
Decays of light mesons are studied by many experimental and theoretical groups, since this allows a deeper understanding of their structure and properties, as well as a better study of the non-perturbative region of QCD. However, semi-leptonic decays of vector mesons are rather poorly studied due to the small values of their branching fractions \cite{ParticleDataGroup:2024cfk}. 

One of the very effective phenomenological models that allows describing the processes of hadron interaction in the non-perturbative energy region (< 2~GeV) is the Nambu-Jona-Lasinio (NJL) model \cite{Nambu:1961tp, Eguchi:1976iz, Ebert:1982pk, Volkov:1984kq, Volkov:1986zb, Ebert:1985kz, Vogl:1991qt, Klevansky:1992qe, Volkov:1993jw, Ebert:1994mf, Hatsuda:1994pi, Buballa:2003qv, Volkov:2005kw}.
     
The main advantage of this model is that it contains a few numbers of parameters (quark masses $m_u \approx m_d$, $m_s$, the cutoff parameter and two four-quark interaction constants) and, as a rule, does not require the introduction of additional arbitrary parameters to describe new types of processes, which increases its predictive power \cite{Volkov:2005kw}. Using this model, in particular, numerous hadronic $\tau$ lepton decays were described in satisfactory agreement with experimental data \cite{Volkov:2017arr, K:2023kgj, Volkov:2023pmy}. 

Semileptonic decays of vector mesons contain vertices similar to these of $\tau$ decays. In this work, in the framework of the NJL model, we study decays containing vertices $\omega \to K \mu \nu_{\mu}$, $\omega \to \pi \mu \nu_{\mu}$, $\omega \to K e \nu_{\mu}$, $\omega \to \pi e \nu_{\mu}$, $\rho \to K \mu \nu_{\mu}$, $\rho \to \pi \mu \nu_{\mu}$, $\rho \to K e \nu_{\mu}$, $\rho \to \pi e \nu_{\mu}$, $K^* \to K \mu \nu_{\mu}$, $K^* \to \pi \mu \nu_{\mu}$, $K^* \to K e \nu_{\mu} $, $K^* \to \pi e \nu_{\mu}$, $\phi \to K \mu \nu_{\mu}$, $\phi \to K e \nu_{\mu}$. Since there are no experimentally measured values for their widths, the obtained results should be considered as predictions. The small number of parameters of the NJL model fixed during its construction makes it a very reliable tool for such predictions. When describing the production of a lepton pair, we proceed from the existence of lepton universality. Namely, the processes of muon and electron production are described by the same coupling constant and only differ in masses. 

\section{Quark-meson Lagrangian of the NJL model}
The fragment of the Lagrangian of the NJL model containing the quark-meson vertices needed for our calculations has the following form \cite{Volkov:1986zb,Volkov:1993jw,Volkov:2005kw}:
\begin{eqnarray}
	\Delta L_{int} & = & \bar{q}\left\{\sum_{i=0,\pm}\left[ig_{\pi}\gamma^{5}\lambda^{\pi}_{i}\pi^{i} +
	ig_{K}\gamma^{5}\lambda^{K}_{i}K^{i} + \frac{g_{\rho}}{2}\gamma^{\mu}\lambda^{\rho}_{i}\rho^{i}_{\mu} + \frac{g_{a_1}}{2}\gamma^{\mu}\gamma^{5}\lambda^{a_1}_{i}a^{i}_{1\mu} + \frac{g_{K^{*}}}{2}\gamma^{\mu}\lambda^{K}_{i}K^{*i}_{\mu}  + \frac{g_{K_1}}{2}\gamma^{\mu}\gamma^{5}\lambda^{K}_{i}K^{i}_{1A\mu}\right] \right. \nonumber\\
	&&\left. + ig_{K}\gamma^{5}\lambda_{0}^{\bar{K}}\bar{K}^{0} + \frac{g_{K^{*}}}{2}\gamma^{\mu}\lambda^{\bar{K}}_{0}\bar{K}^{*0}_{\mu} + \frac{g_{\omega}}{2}\gamma^{\mu}\lambda^{\omega}\omega_{\mu} + \frac{g_{\phi}}{2}\gamma^{\mu}\lambda^{\phi}\phi_{\mu}\right\}q,
\end{eqnarray}
where q is the quark triplet with masses $m_u \approx m_d = 270$~MeV, $m_s = 420$~MeV, $\lambda$ are the linear combinations of the Gell-Mann matrices.

The state $K_{1A}$ is the strange axial vector meson with quantum numbers $J^{PC} = 1^{++}$ splitted into two physical states \cite{Volkov:1984gqw,Suzuki:1993yc, Volkov:2019awd}:
\begin{eqnarray}
    K_{1A} = K_1(1270)\sin{\alpha} + K_1(1400)\cos{\alpha},
\end{eqnarray}
where $\alpha = 57^\circ$.

The coupling constants of the mesons with quarks take the following form:
\begin{displaymath}
	g_{\pi} = \sqrt{\frac{Z_{\pi}}{4 I_{20}}}, \quad g_{\rho} = g_{\omega} = g_{a_1} = \sqrt{\frac{3}{2 I_{20}}}, \quad g_{\phi} = \sqrt{\frac{3}{2 I_{02}}}, \quad g_{K} = \sqrt{\frac{Z_{K}}{4 I_{11}}}, \quad g_{K^{*}} = g_{K_1} = \sqrt{\frac{3}{2 I_{11}}},
\end{displaymath}
where
\begin{eqnarray}
	&Z_{\pi} = \left(1 - 6\frac{m^{2}_{u}}{M^{2}_{a_{1}}}\right)^{-1}, \quad
	Z_{K} = \left(1 - \frac{3}{2}\frac{(m_{u} + m_{s})^{2}}{M^{2}_{K_{1A}}}\right)^{-1},& \nonumber\\
	&M^{2}_{K_{1A}} = \left(\frac{\sin^{2}{\alpha}}{M^{2}_{K_{1}(1270)}} + \frac{\cos^{2}{\alpha}}{M^{2}_{K_{1}(1400)}}\right)^{-1},&
\end{eqnarray}
$Z_{\pi}$ and $Z_{K}$ are the additional renormalization constants appearing as a result of taking into account transitions between axial vector and pseudoscalar mesons.

The integrals $I_{nm}$ appear in quark loops during the renormalization of the Lagrangian and take the following form:
\begin{equation}
\label{integral}
	I_{nm} = -i\frac{N_{c}}{(2\pi)^{4}}\int\frac{\theta(\Lambda^{2} + k^2)}{(m_{u}^{2} - k^2)^{n}(m_{s}^{2} - k^2)^{m}}
	\mathrm{d}^{4}k,
\end{equation}
where $\Lambda = 1265$~MeV is the cut-off parameter \cite{Volkov:2005kw}.

\section{Semileptonic decays of the $\rho$ meson}
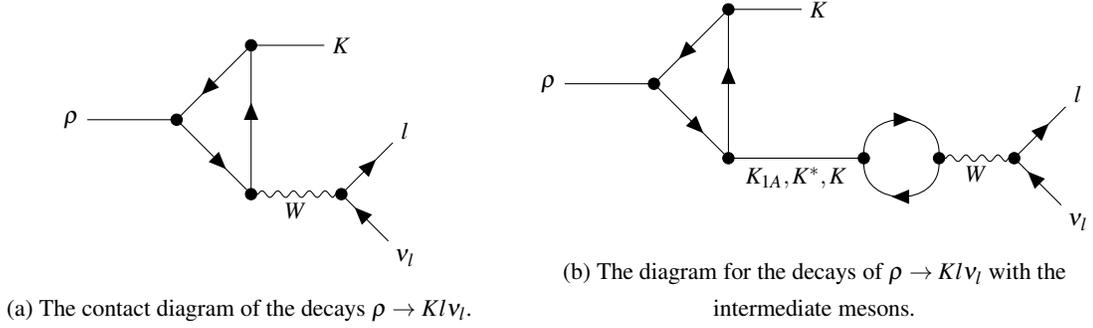
\begin{figure*}[t]
 \centering
  \begin{subfigure}{0.5\textwidth}
   \centering
    \begin{tikzpicture}
     \begin{feynman}
      \vertex (a) {\(\rho\)};
      \vertex [dot, right=1.4cm of a] (b){};
      \vertex [dot, above right=1.4cm of b] (c) {};
      \vertex [dot, below right=1.4cm of b] (e) {};
      \vertex [right=1.2cm of c] (d) {\(K\)};
      \vertex [dot, right=1.2cm of e] (f) {};
      \vertex [above right=1.2cm of f] (g) {\(l\)};
      \vertex [below right=1.2cm of f] (h) {\(\nu_l\)};
      \diagram* {
         (a) -- [] (b),
         (b) -- [anti fermion] (c),
         (c) -- [anti fermion] (e),
         (e) -- [anti fermion] (b),
         (c) -- [] (d),
         (e) -- [boson, edge label'=\(W\)] (f),
         (f) -- [fermion] (g),
         (f) -- [anti fermion] (h),
      };
     \end{feynman}
    \end{tikzpicture}
   \caption{The contact diagram of the decays $\rho \to K l \nu_l$.}
  \end{subfigure}%
 \centering
 \begin{subfigure}{0.5\textwidth}
  \centering
   \begin{tikzpicture}
    \begin{feynman}
      \vertex (a) {\(\rho\)};
      \vertex [dot, right=1.4cm of a] (b){};
      \vertex [dot, above right=1.4cm of b] (c) {};
      \vertex [dot, below right=1.4cm of b] (e) {};
      \vertex [right=1.2cm of c] (d) {\(K\)};
      \vertex [dot, right=1.8cm of e] (f) {};
      \vertex [dot, right=1.0cm of f] (k) {};
      \vertex [dot, right=1.0cm of k] (l) {};
      \vertex [above right=1.2cm of l] (g) {\(l\)};
      \vertex [below right=1.2cm of l] (h) {\(\nu_l\)};
      \diagram* {
         (a) -- [] (b),
         (b) -- [anti fermion] (c),
         (c) -- [anti fermion] (e),
         (e) -- [anti fermion] (b),
         (c) -- [] (d),
         (e) -- [edge label'=\({K_{1A}, K^*, K}\)] (f),
         (f) -- [fermion, inner sep=1pt, half left] (k),
         (k) -- [fermion, inner sep=1pt, half left] (f),
         (k) -- [boson, edge label'=\(W\)] (l),
         (l) -- [fermion] (g),
         (l) -- [anti fermion] (h),
      };
     \end{feynman}
    \end{tikzpicture}
   \caption{The diagram for the decays of $\rho \to K l \nu_l$ with the intermediate mesons.}
  \end{subfigure}%
 \caption{The diagram describing the decays $\rho \to K l \nu_l$.}
 \label{diagram1}
\end{figure*}

In this section, we consider the decays of the $\rho$ meson with breaking and preservation of the strangeness, $\rho^0 \to K^+ l^- \bar{\nu}_{l}$ and $\rho^0 \to \pi^+ l^- \bar{\nu}_{l}$ respectively, where $l=\mu, e$. The diagrams for these decays are presented in Fig. \ref{diagram1}.

The amplitude of the process $\rho^0 \to K^+ \mu^- \bar{\nu}_{\mu}$ in the NJL model takes the following form:
\begin{eqnarray}
\label{amplitude_rho}
\mathcal{M}(\rho^0 \to K^+ \mu^- \bar{\nu}_{\mu}) = \frac{i}{4} G_F V_{us} e_{\mu}(p_{\rho})\left\{T_{ca}^{\mu\nu} + T_{cv}^{\mu\nu} + T_a^{\mu\nu} + T_v^{\mu\nu} + T_p^{\mu\nu} \right\} L_{\nu},
\end{eqnarray}
where $G_{F}$ is the Fermi constant, $V_{us}$ is the element of the Cabibbo-Kobayashi-Maskawa (CKM), $L_\nu$ is the lepton current, $e_\mu(p_{\rho})$ is the polarisation vector of the decaying meson, $T_{ca}^{\mu\nu}$, $T_{cv}^{\mu\nu}$,  $T_a^{\mu\nu}$, $T_v^{\mu\nu}$ and $T_p^{\mu\nu}$ are the contact contributions and axial vector, vector and pseudoscalar channel contributions:
\begin{eqnarray}
\label{channels_rho}
T_{ca}^{\mu\nu} & = & (m_u + m_s)\frac{g_{\rho}}{g_K} Z_K g^{\mu\nu}, \nonumber\\
T_{cv}^{\mu\nu} & = & -i8 m_u g_K g_\rho \left[I_{21} + m_u\left(m_s - m_u\right)I_{31}\right] e^{\mu\nu\lambda\delta} p_{K\lambda} q_{\delta}, \nonumber\\
T_a^{\mu\nu} & = & (m_u + m_s)\frac{g_{\rho}}{g_K} Z_K \left[\sin^2(\alpha) BW_{K_1(1270)} + \cos^2(\alpha) BW_{K_1(1400)}\right] \left\{g^{\mu\nu}\left[q^2 - \frac{3}{2}(m_u + m_s)^2\right] - q^\mu q^\nu\right\}, \nonumber\\
T_v^{\mu\nu} & = & -i8 m_u g_K g_\rho \left[I_{21} + m_u\left(m_s - m_u\right)I_{31}\right] BW_{K^*} \left[q^2 - \frac{3}{2}(m_s - m_u)^2\right]e^{\mu\nu\lambda\delta} p_{K\lambda} q_{\delta}, \nonumber\\
T_p^{\mu\nu} & = & -(m_u + m_s)\frac{g_{\rho}}{g_K} BW_K (p_K - q)^\mu q^\nu,
\end{eqnarray}
where $q$ is the momentum of the intermediate mesons, and $p_K$ is the momentum of the final kaon. The intermediate mesons are described by using the Breit-Wigner propagator
\begin{eqnarray}
\label{BW}
&& BW_{meson} = \frac{1}{M^2_{meson} - q^2 - i \sqrt{q^2} \Gamma_{meson}}, 
\end{eqnarray}
where the masses and widths of the mesons are taken from PDG \cite{ParticleDataGroup:2024cfk}. 

The convergent integrals $I_{21}$ and $I_{31}$ have the same structure as (\ref{integral}).

The decay $\rho^0 \to \pi^+ \mu^- \bar{\nu}_{\mu}$ has a similar structure. The difference is in the preservation of strangeness, and the absence of the vector channel and the appropriate contact contribution. Its amplitude takes the following form:
\begin{eqnarray}
\mathcal{M}(\rho^0 \to \pi^+ \mu^- \bar{\nu}_{\mu}) & = & i G_F V_{ud} e_{\mu}(p_{\rho})\left\{T_{ca}^{\mu\nu} + T_a^{\mu\nu} + T_p^{\mu\nu} \right\} L_{\nu}, \nonumber\\
T_{ca}^{\mu\nu} & = & m_u \frac{g_{\rho}}{g_\pi} Z_\pi g^{\mu\nu}, \nonumber\\
T_a^{\mu\nu} & = & m_u \frac{g_{\rho}}{g_\pi} Z_\pi BW_{a_1} \left\{g^{\mu\nu}\left[q^2 - 6m_u^2\right] - q^\mu q^\nu\right\}, \nonumber\\
T_p^{\mu\nu} & = & - m_u \frac{g_{\rho}}{g_\pi} Z_\pi BW_{\pi} (p_\pi - q)^\mu q^\nu.
\end{eqnarray}

The decays $\rho^0 \to K^+ e^- \bar{\nu}_{e}$ and $\rho^0 \to \pi^+ e^- \bar{\nu}_{e}$ differ from the decays $\rho^0 \to K^+ \mu^- \bar{\nu}_{\mu}$ and $\rho^0 \to \pi^+ \mu^- \bar{\nu}_{\mu}$ only in the mass of the appropriate lepton.

The widths of the decays can be calculated by using the formula

\begin{eqnarray}
\label{width}
\Gamma (\rho \to \pi^0 \mu \nu_\mu) = \frac{1}{3} \cdot \frac{1}{256 {\pi}^3 M^3_{\rho}} \int\limits_{s_{-}}^{s_{+}}ds \int\limits_{t_{-}(s)}^{t_{+}(s)} dt \: {|\mathcal{M}(\rho \to \pi^0 \mu \nu_\mu)|}^2,
\end{eqnarray}
where the Mandelstam variables are defined as $s = (p_{\rho}-p_{\pi})^2=(p_{\mu} + p_{\nu_\mu})^2$, $t=(p_\rho - p_{\nu_\mu})^2=(p_{\pi} + p_{\mu})^2$. The integration limits take the form
\begin{eqnarray}
s_{+} = (M_{\rho} - M_{\pi})^2, \quad s_{-} = M^2_{\mu},
\end{eqnarray}
 
\begin{eqnarray}
\label{L1}
&& t_{\pm}(s) = \frac{1}{2} \biggl[ M^2_{\rho} + M^2_{\pi} + M^2_{\mu} - s + \frac{M^2_\mu(M^2_{\rho}-M^2_\pi)}{s} \pm \sqrt{ s^{-2} \cdot \Omega(s)} \biggr],
\end{eqnarray}
where $\Omega(s)=(M^2_\mu - s)^2 \cdot (M^4_{\rho}+(M^2_\pi - s)^2-2M^2_{\rho}(M^2_\pi+s))$. 

The results for the widths of the $\rho$ meson decays are given in Table~\ref{tab_width}.

\section{Semileptonic decays of $\omega$ and $\phi$ mesons}
The decay $\omega \to K^+ \mu^- \nu_{\mu}$ proceeds with the breaking of strangeness and is described by the amplitude of the same structure as the appropriate amplitude for the $\rho$ meson (\ref{amplitude_rho}).

The amplitude of the decay $\omega \to K \mu \nu_{\mu}$ proceeding with the preservation of strangeness differs from the amplitude of the $\rho$ meson decay in that it contains only the vector channel and the corresponding contact contribution:
\begin{eqnarray} 
\mathcal{M}(\omega \to \pi^+ \mu^- \bar{\nu}_{\mu}) & = &  \frac{3}{8\pi} G_F V_{ud} \frac{g_\rho g_\pi}{m_u} \left[1 + q^2 BW_{\rho}\right] e_{\mu}(p_{\omega})e^{\mu\nu\lambda\delta} p_{\pi\lambda} q_{\delta} L_{\nu}.
\end{eqnarray}

The $\phi$ meson decay of this type takes place only with breaking of the strangeness $\phi \to K \mu \nu_{\mu}$. Its amplitude can be presented in the following form:
\begin{eqnarray}
\mathcal{M}(\phi \to K^+ \mu^- \bar{\nu}_{\mu}) & = & i\frac{\sqrt{2}}{4} G_F V_{us} e_{\mu}(p_{\rho})\left\{T_{ca}^{\mu\nu} + T_{cv}^{\mu\nu} + T_a^{\mu\nu} + T_v^{\mu\nu} + T_p^{\mu\nu} \right\} L_{\nu}, \nonumber\\
T_{ca}^{\mu\nu} & = & (m_u + m_s)\frac{g_{\phi}}{g_K} Z_K g^{\mu\nu}, \nonumber\\
T_{cv}^{\mu\nu} & = & i 8 m_s g_K g_\phi \left[I_{12} - m_s\left(m_s - m_u\right)I_{13}\right] e^{\mu\nu\lambda\delta} p_{K\lambda} q_{\delta}, \nonumber\\
T_a^{\mu\nu} & = & (m_u + m_s)\frac{g_{\phi}}{g_K} Z_K \left[\sin^2(\alpha) BW_{K_1(1270)} + \cos^2(\alpha) BW_{K_1(1400)}\right] \left\{g^{\mu\nu}\left[q^2 - \frac{3}{2}(m_u + m_s)^2\right] - q^\mu q^\nu\right\}, \nonumber\\
T_v^{\mu\nu} & = & i 8 m_s g_K g_\phi \left[I_{12} - m_s\left(m_s - m_u\right)I_{13}\right] BW_{K^*} \left[q^2 - \frac{3}{2}(m_s - m_u)^2\right]e^{\mu\nu\lambda\delta} p_{K\lambda} q_{\delta}, \nonumber\\
T_p^{\mu\nu} & = & -(m_u + m_s)\frac{g_{\phi}}{g_K} BW_K (p_K - q)^\mu q^\nu.
\end{eqnarray}

The widths of the decays of the $\omega$ and $\phi$ mesons are presented in the Table~\ref{tab_width}.

\section{Description of semileptonic decays of the meson $K^*$}  
We consider semileptonic decays of the strange vector meson $K^*$ proceeding with a weak $s-u$ transitions $K^* \to \pi^0 l \bar{\nu_l}$ and decay $K^* \to \bar{K^0} l \bar{\nu_l}$, where $l=\mu, e$. 
The decay $K^* \to \pi^0 l \bar{\nu_l}$ is described by a diagram with contact interaction with the vertex $K^* \to \pi^0W$ containing the transition $W\to l \nu_l$ and a diagram with intermediate mesons $K_{1A}$, $K^*$ and $K$.

Calculations within the NJL model give the following amplitude for the semileptonic decay of the strange vector meson $K^* \to \pi^0 \mu \bar{\nu_\mu}$:
	\begin{eqnarray}
	\label{amplitude_1}
\mathcal{M}(K^* \to \pi^0 \mu \nu_\mu) & = & i G_F V_{us} e_\mu \left[ T^{\mu\nu}_a + T^{\mu\nu}_v + T^{\mu\nu}_p
\right] L_\nu,
	\end{eqnarray}
where separate contributions take the form 
\begin{eqnarray}
T^{\mu\nu}_a & = & 3 m_s \frac{g_\pi}{g_{K^*}} \left[g_{\mu\nu} + \biggl[ g_{\mu\nu} \left( q^2 - \frac{3}{2} (m_s+m_u)^2 \right) - q_\mu q_\nu \biggl] BW_{K_1(1270)} \sin^2(\alpha)  \right. \nonumber\\
        && \left. + \biggl[ g_{\mu\nu} \left( q^2 - \frac{3}{2} (m_s+m_u)^2 \right) - q_\mu q_\nu \biggl] BW_{K_1(1400)} \cos^2(\alpha) 
\right],
\end{eqnarray}
\begin{eqnarray}
T^{\mu\nu}_v & = & 2 m_u g_{K^*} g_\pi \left( I_{21} + m_u (m_s-m_u) I_{31}\right) \left[1 +\left( q^2 - \frac{3}{2} (m_s-m_u)^2 \right) BW_{K^*} 
\right] \epsilon_{\mu\nu\lambda\delta} p_{\pi\lambda} q_{\delta},
\end{eqnarray}
\begin{eqnarray}
T^{\mu\nu}_p & = & - \frac{3(m_s+m_u) g_\pi}{g_{K^*}} BW_{K} {\left(p_{\pi}-q \right)}_\mu q_\nu,
\end{eqnarray}
where $e_\mu$ is the polarization vector of the meson $K^{*}$. 

The decay $K^* \to \bar{K^0} \mu \bar{\nu_\mu}$ is described by the following amplitude:
	\begin{eqnarray}
	\label{amplitude_2}
\mathcal{M}(K^* \to \pi^0 \mu \nu_\mu) & = & i \sqrt{2} G_F V_{ud} e_\mu \left[ T^{\mu\nu}_a + T^{\mu\nu}_v + T^{\mu\nu}_p
\right] L_\nu,
	\end{eqnarray}
where
	\begin{eqnarray}
T^{\mu\nu}_a & = &  \frac{ Z_K (3m_u-m_s)}{4} \frac{g_{K^*}}{g_K} \left[g_{\mu\nu} + \biggl( g_{\mu\nu} \left( q^2 - 6m^2_u \right) - q_\mu q_\nu \biggl) BW_{a_1} 
\right],
	\end{eqnarray}
\begin{eqnarray}
T^{\mu\nu}_v & = & 2 m_u g_{K^*} g_K \left( I_{12} - m_u (m_s-m_u) I_{13}\right) \left[1 + q^2 BW_{\rho} 
\right] \epsilon_{\mu\nu\lambda\delta} p_{K\lambda} p_{\delta},
\end{eqnarray}
 \begin{eqnarray}
T^{\mu\nu}_p & = & - \frac{m_u Z_K g_{K^*}}{g_K} BW_{\pi} {\left(p_K-q \right)}_\mu q_\nu.
\end{eqnarray}

The amplitudes for the $e\bar{\nu_e}$ lepton pair production processes are obtained by replacing the mass $M_\mu \to M_e$.
The widths of the semileptonic decays of the meson $K^*$ calculated using the amplitudes \ref{amplitude_1} and \ref{amplitude_2} are given in Table \ref{tab_width}.

\begin{table}[h!]
\begin{center}
\begin{tabular}{ccc}
\hline
Decays &  $\Gamma_{\mu \nu_\mu}$ & $\Gamma_{e \nu_e}$  \\
\hline
$\rho \to K l \nu_l$ 		        & 11.5	& 187.2 \\
$\rho \to \pi l \nu_l$ 		        & 588.7 & 2012.4 \\
$\omega \to K l \nu_l$		        & 1.54  & 23.6 \\
$\omega \to \pi l \nu_l$		    & 5.91  & 6.68  \\
$\phi \to K l \nu_l$		        & 222.3 & 3330.5\\
$K^* \to \bar{K^0} l \bar{\nu_l}$   & 2.96	    & 4.10   \\
$K^* \to \pi l \nu_l$		        & 3.52	    & 3.82   \\
\hline
\end{tabular}
\end{center}
\caption{Semileptonic decay widths of vector mesons in $\Gamma \times 10^{14}$ MeV}
\label{tab_width}
\end{table} 

\section{Conclusion}
The calculations of weak semileptonic decays of vector mesons within the NJL model lead to small decay widths, which explains the absence of experimental data in this area. However, there are experimental data for the decays $K \to \pi \mu \nu_\mu$ and $K \to \pi e \nu_e$ which are close in structure to the processes considered here. The branching fractions of these decays turn out to be much more attainable for experimental observations: $Br(K \to \pi \mu \nu_\mu)=(3.35 \pm 0.03)\%$ and $Br(K \to \pi e \nu_e)=(5.07 \pm 0.04)\%$ \cite{ParticleDataGroup:2024cfk}. It is interesting to note that the absolute widths of these decays are two orders of magnitude lower than the decay widths of the processes considered here: $\Gamma(K \to \pi \mu \nu_\mu) = (1.871 \pm 0.018) \times 10^{-15}$ MeV and $\Gamma(K \to \pi e \nu_e) = (2.695 \pm 0.021) \times 10^{-15}$ MeV \cite{ParticleDataGroup:2024cfk}.

Theoretical estimations of the decays $K \to \pi l \nu_l$ were made in \cite{Leutwyler:1984je}. The estimations within the NJL model also lead to agreement with the experiments $Br(K \to \pi \mu \nu_\mu)_{NJL} = (2.96 \pm 0.44)\%$ (see Appendix \ref{appendix}). This allows us to hope for the reliability of the obtained predictions for weak semileptonic decays of vector mesons.

The obtained results for the decay widths show that very high accuracy is necessary for planned experiments. Unfortunately, such accuracy in measuring vector meson decays has not yet been achieved at modern facilities.
If future experiments lead to higher data for branching fractions, this could be considered as an indication of the manifestation of effects beyond the Standard Model.

\appendix
\section{The decays $K \to \pi \mu \nu_\mu (e \nu_e)$}
\label{appendix}
The semileptonic decays of the kaon with production of pion and lepton pairs $\mu \nu_\mu$ and $e \nu_e$ are described by the contribution of the contact diagram and the channel with the intermediate strange vector meson $K^*$. These channels correspond to similar diagrams presented in Figure \ref{diagram1}. The total amplitude of the decay $K \to \pi \mu \nu_\mu (e \nu_e)$ takes the form
	\begin{eqnarray}
	\label{amplitude_3}
\mathcal{M}(K \to \pi \mu \nu_\mu) & = &  \frac{3 g_K g_\pi}{g^2_{K^*}}G_F V_{us} \left( A_K p_{K\mu} + p_{\pi\mu} \right) 
\left[g_{\mu\nu} + \biggl( g_{\mu\nu}f(q^2) - q_\mu q_\nu f(M^2_{K^*}) \biggl) BW_{K^*} \right] L_\nu,
	\end{eqnarray}
where $f(q^2) = 1 - 3(m_s-m_u)^2/2q^2$; $q=p_\mu + p_{\nu_\mu}$; $p_{K}$, $p_{\pi}$ are meson momenta. The factor $A_K$ appears as a result of taking into account the non-diagonal transitions between the axial-vector and pseudoscalar mesons in the external line
	\begin{eqnarray}
        A_K = 1- \frac{3 m_s (m_s+m_u)}{M^2_{K_{1A}}}.  
	\end{eqnarray}
   
As a result taking into account the $K_1 - K$ transition, we obtain $Br(K \to \pi \mu \nu_\mu)_{NJL} = (2.96 \pm 0.44)\%$ for the decay width. The uncertainty of the model predictions for the decay widths is estimated at the level of $15\%$ \cite{Volkov:2005kw,Volkov:2017arr}.
The calculated decay widths agree with the experimental data: $Br(K \to \pi \mu \nu_\mu)=(3.35 \pm 0.03)\%$.
Taking into account the non-diagonal $K_1-K$ transition is justified by the results for the strong decay of $K^* \to K \pi$ where the theoretical estimate of the width $\Gamma(K^* \to K \pi)_{NJL} = 58.0 \pm 8.85$ MeV at the experimental value $\Gamma(K^* \to K \pi)_{exp} = 51.4 \pm 0.8$ MeV \cite{ParticleDataGroup:2024cfk}.

\subsection*{Acknowledgments}
The authors are grateful to A.~B.~Arbuzov for useful discussions. 
This work was supported by the Science Committee of the Ministry of Science and Higher Education of the
Republic of Kazakhstan Grant no. BR21881941.

\end{document}